# Real-time discrete multi-tone transmission for passive optical networks in C- and O-band

*Annika Dochhan[1], Tomislav Drenski[2], Helmut Griesser[1], Michael H. Eiselt[1], Jörg-Peter Elbers[1]*

[1]ADVA Optical Networking SE, Märzenquelle 1-3, 98618 Meiningen, Germany
[2]Socionext Europe GmbH, Concorde Park, Concorde Road, SL64FJ, UK

**Abstract**

DMT at 25 Gbit/s, 50 Gbit/s and 100 Gbit/s in passive systems with electro-absorption modulators is investigated. Transmission in C- and O-band are compared, yielding up to 22 km reach, 100 Gbit/s in O-band. All results are obtained with a real-time signal processing ASIC.

## 1  Introduction

The upcoming introduction of 5G will impose new challenges onto passive optical networks (PON) in terms of required data rates. For fronthaul networks, several options are in discussion, including wavelength division multiplexing (WDM-PON), as in ITU-T Study Group 15 recommendation G.698.4 [1]. In this case, each wavelength is carrying 10 Gbit/s of data with a potential upgrade to 25 Gbit/s. Another potential approach is based on Ethernet system daisy chain cell site network interface devices with one or few 100G trunk lines leveraging statistical multiplexing [2]. This approach would be based on a system with one high data rate interface, e.g. 100 Gbit/s and cheap grey optics (e.g. in O-band) and offers an upgrade path to coarse WDM. This data rate requires higher order modulation to reduce the bandwidth requirements of the components, while for WDM-PON, currently binary non-return to-zero (NRZ) modulation is considered. However, for future higher rates, also WDM-PON will require higher order modulation. Besides four-level pulse amplitude modulation (PAM4) which is heavily discussed to meet these requirements, discrete multi-tone modulation (DMT) has also been proposed, offering similar performance in terms of noise sensitivity and higher chromatic dispersion (CD) tolerance [3-5].

In this work, DMT is investigated with the first available real-time digital signal processing chip for various data rates, covering 25 Gbit/s, 50 Gbit/s and 100 Gbit/s. To meet the requirements of low cost, purely passive transmission is performed, without any optical amplification. Two electro-absorption modulators integrated with distributed feedback lasers (DFB/EAM = EML) are used, which are available with a similar design for C- and O-band. It is evident that, for higher data rates, the reach is mainly limited by CD, hence in the O-band up to 22 km using standard single mode fibre (SSMF) can be bridged even with 100 Gbit/s, while for lower rates, the available power is the only limiting factor.

## 2.  Experimental Setup

The experimental setup is shown in Fig.1. The Socionext real-time DMT DSP ASIC chip, which was mounted on an evaluation board, was connected to a differential input linear electrical driver with 30 GHz bandwidth. A subsequent bias-tee is used to couple the signal with an optimized DC bias (-1.2 V for O-band EML and -1.05 V for C-band EML) to a 25-GHz bandwidth EML. The EML is a ridge-waveguide based InGaAlAs component, designed by Fraunhofer Heinrich-Hertz-Institute [6-7], and has the potential of being a low-cost product for the targeted applications discussed in this paper. The design is similar for C- and O-band, but the C-band device has less output power. The devices were operated at a laser current of 100 mA, which led to a wavelength of 1309 nm for the O-band device and 1565.4 nm for the C-band laser. This is at the upper end of the C-band, therefore, the results shown can be considered as worst case reference for the evaluation of CD influences, which will be smaller for shorter wavelengths. After the EML, an isolator limits any back reflection of the light. The signal is transmitted over up to 32.6 km of SSMF, using always the maximum possible fibre launch power (2.5 dBm for O-band, -2 dBm for C-band). The input to the receiver can be varied by a variable optical attenuator (VOA). The receiver is a 35-GHz bandwidth PIN photodiode in combination with a linear transimpedance amplifier (TIA). The differential output of this amplifier is connected to the DMT chip.

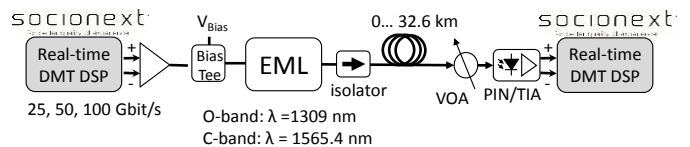

Fig. 1 System setup for passive discrete multi-tone transmission with electro absorption modulator (EML) in O- and C-band at 25, 50 and 100 Gbit/s.

The DMT chip runs at ~64 GS/s for 100 Gbit/s and 50 Gbit/s and at ~42 GS/s for 25 Gbit/s. Other data rates could be chosen/programmed as well. The signal consists of 256 subcarriers, where some are used as pilots for synchronization and equalization purposes. A cyclic prefix of 16 samples is used and the signal RMS (root mean square, related to clipping ratio) is dynamically optimized during the measurement. The net data rates correspond to the Ethernet data rates, i. e. 103.125 Gbit/s for the 100G mode, and half/quarter of that for 50G/25G mode. For simplicity, we note them as 25, 50 and 100 Gbit/s throughout paper. Forward error correction (FEC) is implemented in the chip. This CI-BCH FEC has a bit error ratio (BER) limit of 4.4e-3. The FEC is working stable during the measurement, giving post-FEC error free results at a BER close to the limit (above 3e-3) for all long term measurements (over night) performed. Due to time limitations during the measurements, longer than one night was not tested, but since there is a significant temperature difference in the lab between night and day, which can also influence the optical setup, these measurements also show the good performance of the adaptive equalization. The chip has implemented bit and power loading (BL, PL) and adaptive back ground equalization to adopt to slowly varying channels. The best BL and PL is found with a water-filling algorithm, using the signal-to-noise ratio (SNR) information of the channel, which is estimated with dedicated pilot tones.

## 3. Results and Discussion

First, the sensitivity of the modulation is determined for each data rate and both EMLs without any fibre. The results are shown in Fig. 2 and Fig. 3. For the O-band, the input power to the Rx could exceed the optimum of -3 dBm (also depending on TIA gain settings, where optimization is not shown here), while for the C-band, the received power is limited to this value, since the VOA adds 1 dB of insertion loss even at minimum attenuation. For 25 Gbit/s, the C-band EML shows slightly higher sensitivity, while for 100 Gbit/s the O-band EML works slightly better. In the O-band, above -7 dBm, and in the C-band, above -6 dBm, 25 Gbit/s could be detected error free. The estimated SNR, which is a good measure for the capacity of the channel, for 64 GS/s probing at $P_{rec}= -3$ dBm (i.e. for 50 Gbit/s or 100 Gbit/s) is depicted in Fig. 4 (a) and (b) in red. The SNR at low frequencies is slightly lower for the C-band than for the O-band. This could be attributed to a different bias point which might lead to stronger modulation non-linearities from the modulator. This effect would be seen in the SNR as additional noise.

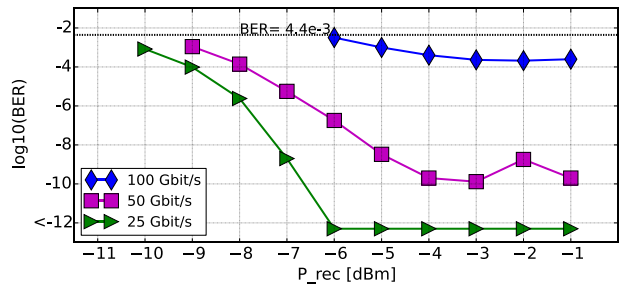

Fig. 2 BER vs. received power for 1309 nm EML.

In the next step, fibre is added to the setup. In the O-band no significant influence from the fibre on the signal can be seen as long as the optimum received power of -3 dBm is maintained, as can be concluded from Fig. 5. At 16 km, the maximum received power was -4.3 dBm, therefore the BER started to degrade. The estimated SNRs with maximum transmission reach for 100 und 50 Gbit/s are shown in Fig. 4 (e) and (f). The degradation of the SNR is clearly visible, but no additional influence like CD or chirp can be seen. In addition to the estimated SNR, the bit loading is displayed (see (c) and (g)). There are only minor differences between the back-to-back and the transmission case. 100 Gbit/s could be transmitted over up to 22.4 km, while for both, 50 Gbit/s and 25 Gbit/s 32.6 km could be achieved.

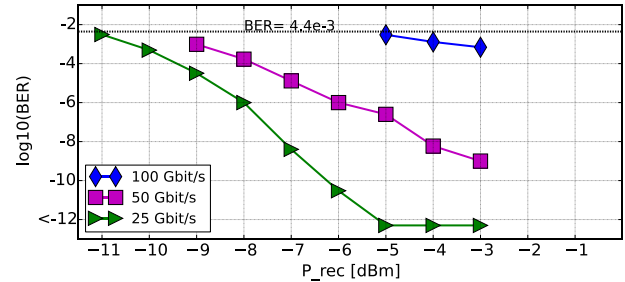

Fig. 3 BER vs. received power for 1565.4 nm EML.

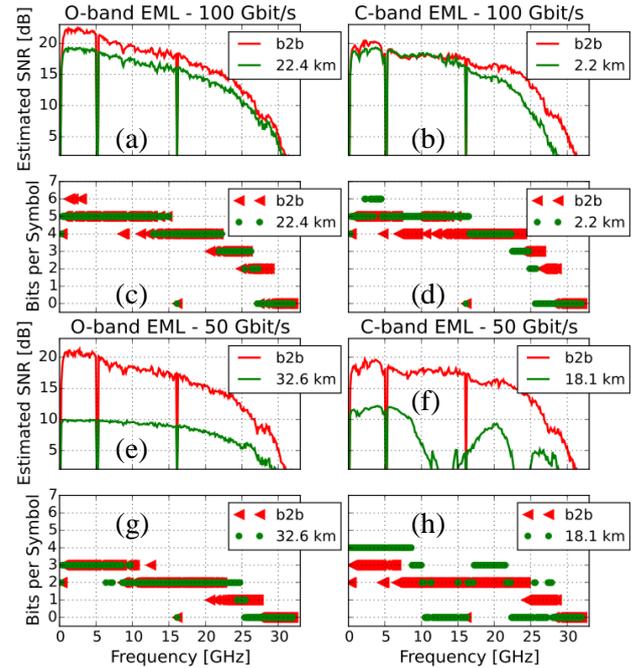

Fig. 4 Estimated SNR vs. frequency (a, b, e, f) and resulting bit loading (c, d, g, h) for both EMLs at 100 Gbit/s and 50 Gbit/s. Red: Back-to-back, Green: after maximum transmission reach.

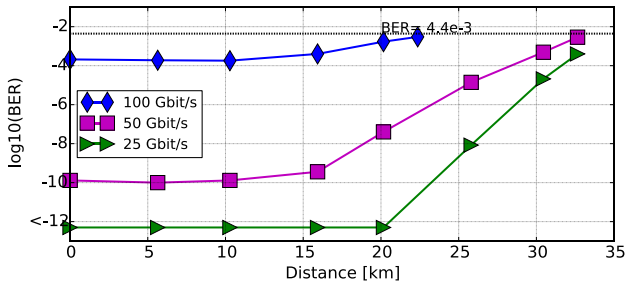

Fig. 5 BER vs. transmission distance for 1309 nm EML.

In the case of the C-band EML, the effect of CD already reduced the performance drastically for 100 Gbit/s at short distances. Fig. 4 (b) shows this bandwidth reduction for 2.2 km of transmission, which was the longest possible distance in the C-band. The whole available bandwidth of the electrical and electro-optical components was needed to transmit the 100 Gbit/s. Therefore, already the small bandwidth reduction from the first CD notch in the spectrum has a severe impact on the signal. The BER vs. distance is shown in Fig. 6. In an amplified system, the performance in the C-band typically is drastically reduced after the first few km, when the first CD notch enters the spectrum. From that point on, the performance only slowly reduces, since the notches shift closer together and thus the gaps which cannot carry any data are narrower [8]. However, in a system without amplifiers, the received power decreases with distance, and this effect is superimposed on the CD effect. This can be clearly seen from the curve for 50 Gbit/s in Fig. 6 – the strong BER decrease between 0 and 5 km results from CD mostly, while for the following distances, the reduced input power is the worsening effect. The estimated SNRs and the corresponding bit loading for the maximum reaches are shown in Fig.4 (b), (d), (f) and (h). The CD notches for 50 Gbit/s require the increase of the modulation order of many subcarriers. In the C-band, the reach is limited to 2.2 km for 100 Gbit/s, to 18.1 km for 50 Gbit/s and to 30.4 km for 25 Gbit/s for this un-amplified system/test setup. This result clearly shows that if sufficient bandwidth is available, DMT can handle CD effectively. The reach for the C-band EML is only few km shorter than for the O-band EML. In both cases, the received power was approximately -10 dBm, therefore the limit was only the power.

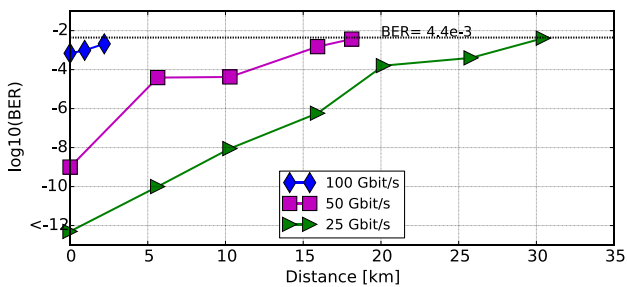

Fig. 6 BER vs. transmission distance for 1565.4 nm EML.

## 4. Conclusion

Discrete multi-tone transmission at 25 Gbit/s, 50 Gbit/s and 100 Gbit/s was evaluated, using EMLs in O- and C-band in an un-amplified system. In O-band the system is only power limited, leading to a reach of 22.4 km for 100 Gbit/s and 32.6 km for both, 50 and 25 Gbit/s. In C-band, chromatic dispersion limits the reach at 100 Gbit/s to 2.2 km, while 50 Gbit/s still allows 18.1 km. For 25 Gbit/s, chromatic dispersion has no significant impact, since the bandwidth allows the DMT bit- and power loading to compensate for it. All results were obtained in real-time with a fully functional DMT ASIC chip, including forward error correction running stable in long-term over-night measurements. The results show that DMT at 25 Gbit/s could serve well as an upgrade for future WDM-PON with C-band modulators. For fronthaul scenarios with 100 Gbit/s and grey or coarse WDM systems, the O-band would be the right choice.

## 5. Acknowledgement

The results were obtained in the projects SENDATE Secure-DCI and SpeeD, partly funded by the German ministry of education and research (BMBF) under contracts 16KIS0477K and 13N1374, respectively.